\documentclass[iop]{emulateapj}

\usepackage{amsmath, amsthm, amssymb, slashed}
\usepackage{color}
\usepackage{graphicx}
\usepackage[toc]{appendix}
\usepackage{times}
\usepackage{upgreek}
\usepackage{hyperref}

\newcommand{\vp}{\varphi}
\newcommand{\dd}{\, {\rm d}}
\newcommand{\p}{^{\prime}}

\newcommand{\rF}{r_{\rm F}}
\newcommand{\rp}{r_{\rm p}}
\newcommand{\rxc}{r_{\rm X}^{\rm c}}

\newcommand{\rx}{r_{\rm X}}
\newcommand{\wx}{w_{\rm X}}
\newcommand{\Tet}{\Theta_{\rm X}}
\newcommand{\Tetn}{\Theta_{\rm X}^0}
\newcommand{\pdv}[2]{\frac{\partial #1}{\partial #2}}
\newcommand{\ralp}{r_{\alpha}}
\newcommand{\rbet}{r_{\beta}}
\newcommand{\zs}{z_{\rm s}}
\newcommand{\rs}{r_{\rm s}}
\newcommand{\rsc}{r_{\rm s}^{\rm c}}
\renewcommand{\vec}{\mathbf}
\begin{document}

\title{Solenoidal improvements for the JF12 Galactic magnetic field model}

\author{Jens Kleimann}
\email{jk@tp4.rub.de}
\author{Timo Schorlepp} 
\email{timo.schorlepp@rub.de}
\author{Lukas Merten}
\email{lukas.merten@rub.de}
\author{Julia Becker Tjus}
\email{julia@tp4.rub.de}

\affil{Ruhr-Universit\"at Bochum, Fakult\"at f\"ur Physik und Astronomie,
       Institut f\"ur Theoretische Physik IV, 44780 Bochum, Germany}

\begin{abstract}
The popular JF12 analytic model by Jansson \& Farrar (2012) provides a
quantitative description of the Galaxy's large-scale magnetic field, which
is widely used in various astrophysical applications. However, both the
poloidal X-type component and the spiral disk component of JF12 exhibit
regions in which the magnetic divergence constraint is violated.
We first propose a cure for this problem, resulting in a truly solenoidal
large-scale spiral field. Second, the otherwise straight field lines of the X-type
component exhibit kinks in the Galactic plane that, in addition to implying
the presence of a singular current sheet,
may pose difficulties for e.g., numerical tracing of cosmic-ray particles.
We propose and discuss two possible strategies to mitigate this problem.
Although all corrections are kept as minimal as possible, the extended set of
model parameters will have to be carefully readjusted in order to fully restore
the agreement to observational data that the unmodified JF12 field is based on.
Furthermore, the performance of our improved version of the field model is
quantitatively assessed by test simulations using the CRPropa Galactic
cosmic-ray propagation code. \\
\end{abstract}


\section{Introduction}

Knowing the structure of the magnetic field of the Milky Way is crucial for various applications,
such as the understanding of cosmic-ray transport in the Galaxy or magnetohydrodynamic (MHD)
models of Galactic evolution. Only a full description of both the turbulent and the regular
components of the field enables the prediction of cosmic-ray signatures at Earth under
consideration of a realistic diffusion tensor.  One of the more recent approaches to a full
description on the basis of analytic equations was developed by 
\citet[][hereafter JF12]{Jansson_Farrar:2012}.
An improved version of the JF12 model was developed and applied -- although not described in full
detail -- more recently by \citet{Unger_Farrar:2017, Unger_Farrar:2019}. As of today, the JF12
field and its updates represent the most complete analytic description of the global Galactic
magnetic field (GMF).

The ``regular,'' large-scale part of the JF12 model comprises four field components: a spiral disk
field, a molecular ring inside the spiral field, a toroidal halo field, and a poloidal  X-shaped
field. An additional ``striated'' component is parameterized as scaling uniformly with the regular
field. \citet{Jansson_Farrar:2012}, who were the first to include an X-shaped component into their
model, also rightfully stressed the importance of obeying Maxwell's magnetic solenoidality
constraint
\begin{equation}
  \label{eq:divB0}
  \nabla \cdot \vec{B} = 0
\end{equation}
but, upon closer inspection, do not meet the latter requirement at all boundaries. An alternative
family of GMF models fully satisfying Equation~(\ref{eq:divB0}) was
developed by \citet{Ferriere_Terral:2014} and later adjusted to constraints from observational
data by \citet{Terral_Ferriere:2017}. More recently, \citet{Shukurov_EA:2019} presented a
parameterized GMF model based on magnetic diffusion and mean-field dynamo theory.
Unlike the upcoming IMAGINE project \citep{Boulanger_EA:2018}, which aims at the development of a
completely revised GMF model by combining current observational data from various sources with
modern Bayesian analysis, this work does not present a completely new model. Instead, we suggest a
gradual improvement of the existing JF12 model, although the employed ideas might well be used in
future GMF models.

The paper is organized as follows. After this present introduction, Section~\ref{sec:spiral_field}
describes and discusses two methods to turn JF12's spiral field component into a truly
solenoidal version of itself. A much simpler solution is then offered for a similar problem with
the X-type field component in Section~\ref{sec:x-type_field}, which then presents two possible
modifications to remove the sharp kinks of field lines at the Galactic plane while still
maintaining solenoidality. Section~\ref{sec:performance} contains a performance comparison of the
old vs.\ the new field model using simulations of propagating particles, and
Section~\ref{sec:summary} concludes the paper with a summary.
Throughout this paper, only the spiral disk and the X-field will be addressed. Neither the
molecular ring, nor the toroidal halo field or the two turbulent components are subject to
investigation in this paper.
\\

\section{A Truly Solenoidal Spiral Disk}
\label{sec:spiral_field}

\subsection{Motivation}

We argue that it is vital for a GMF model to be completely void of magnetic monopoles both
globally and locally at least for the following two reasons.
First, depending on the application at hand and the methods employed therein, even small nonzero
values of $|\nabla \cdot \vec{B}|$ may give rise to unphysical effects such as negative pressures
or densities, violation of momentum or energy conservation, or the rise of spurious waves
\citep[e.g.][]{Brackbill_Barnes:1980}, specifically in the context of MHD simulations, including
cases where the magnetic field is not actually evolved but treated as a static background.

Second, due to our fixed position within the Galaxy and the fact that only line-of-sight
observations from this solitary vantage point are available, our Galaxy's global properties (such
as shape, geometry, and magnetic field structure, to name but a few) are inherently difficult to
constrain. Intrinsic ambiguities have to be resolved through inversion and parameter fitting.
It is therefore all the more important to use as many physical constraints as possible. In that
vein, the unconditional validity of Maxwell's equations, and in particular of
Equation~(\ref{eq:divB0}), is clearly undisputed and provides rather tight constraints on the set
of physically admissible field models, as was already noted by \citet{Jansson_Farrar:2012}.
The effect is expected to be strongest for studies sensitive to spurious magnetic monopoles.
Even in other cases, the use of a completely solenoidal field model is to be preferred for its
higher degree of physical realism. In particular, it is important to keep in mind that violations
of Equation~(\ref{eq:divB0}), even those that are limited to a spatial volume of measure zero,
will often have far-reaching consequences also for more distant regions. This latter point will be
illustrated in this paper on the basis of the marked differences between GMF models that conserve
magnetic flux and those that do not. \\

\subsection{Properties of the JF12 Spiral Disk Field}

We begin by briefly summarizing the basic properties of the original JF12 spiral disk component,
taking the opportunity to properly write up the relevant equations. In the disk region between
$r_1=5$ and $r_2=20$~kpc, a field line passing through a point with supergalactic cylindrical
coordinates $(r_{\rm a}, \vp_{\rm a}, z_{\rm a})$ follows a logarithmic spiral
\begin{equation}
  r(\vp) = r_{\rm a} \exp [ (\vp-\vp_{\rm a}) \tan i ]
\end{equation}
with a uniform inclination angle $i=11.5^{\circ}$, as depicted in the left panel of
Figure~\ref{fig:compare_diskfield}. The spiral is partitioned into eight field line-delimited
regions of relative widths $f_j$ summing to $\sum_{j=1}^8 f_j = 1$ (actually to $0.999$ due to
round-off errors), with corresponding field strengths $b_j$ at the inner rim $r_1$.
The field strength parameters $b_{1...7}$ are fitted to data, while $b_8$ is chosen such that
\begin{equation}
  \label{eq:normflux}
  \sum_{j=1}^8 b_j \, f_j = 0
\end{equation}
holds, implying that the total magnetic flux passing through a coaxial cylindrical shell of any
radius is zero. The field strength in the spiral region $j$ is equal to $(r_1/r) \, b_j$. While this,
together with constraint (\ref{eq:normflux}), is sufficient to warrant magnetic solenoidality
within the disk, field lines will still ``start'' and ``end'' at the inner and outer spiral disk
boundaries, i.e., Maxwell's divergence constraint (\ref{eq:divB0}) is violated along
these boundaries, despite occasional claims to the contrary
\citep[e.g.][]{Jansson_Farrar:2012, Beck_EA:2016, Unger_Farrar:2019}.
We understand the spiral field's confinement to the annulus \mbox{$r \in [r_1,r_2]$} to be
motivated by the fact that the underlying data analysis would not allow the field to be
adequately constrained beyond these radii, implying the need to substitute a suitable solenoidal
continuation before employing the field in a particular scientific investigation. In this vein,
we propose and discuss possible solutions to this crucial issue in Section~\ref{sect:fix_spirals}.

\subsection{Explicit Component Formulas}

Since JF12 provide explicit formulas only to some extent, and partially content themselves with
mere recipes for the construction of the actual field components, we use the opportunity to
provide these formulas here for completeness and later reference, and in a form that will be more
suitable for the purpose at hand.

JF12 specify the border between adjacent spiral regions by means of the radius $r_{-x}$ at which a
spiral boundary intersects the negative $x$-axis. We note in passing that their spiral equation
$r = r_{-x} \exp [\vp \tan(90^{\circ} - i)]$ should actually read
$r = r_{-x} \exp [(\vp -\pi) \tan i]$ as it would otherwise relate $r_{-x}$ to the $(\vp=0)$
direction, i.e., the positive $x$-axis, and would furthermore result in a much larger
inclination angle of \mbox{$90^{\circ}-i = 78.5^{\circ}$} that would have field lines pointing
outward almost radially.

This functional form and the value for $i$ were adopted from the earlier model by
\citet{Brown_EA:2007}, although these authors do not cite explicit values for $r_{-x}$.
For our purpose, and possibly for other applications as well, it is instead more convenient to
work in terms of the azimuthal angle $\vp=\Phi_{1,j}$ at which the limiting field line $r_j(\vp)$
between two adjacent regions, $j$ and $j-1$, intersects the inner spiral disk boundary at
$r_1 = 5$~kpc. (For the remainder of this paper, all lengths are in units of kpc unless indicated
otherwise.) These two descriptions are related through
\begin{equation}
  \frac{r_{-x,j}}{r_1} = \frac{r_j(\pi)}{r_j(\Phi_{1,j})} =
  \exp \left[ (\pi-\Phi_{1,j}) \tan i \right] ,
\end{equation}
and the relative width of spiral region $j$ is
\begin{equation}
  f_j = (\Phi_{1,j-1}-\Phi_{1,j})/(2\pi)
\end{equation}
with $\Phi_{1,0}\equiv 2\pi+\Phi_{1,8}$ for cyclic closure.
Table~\ref{tab:widths} summarizes the obtained values. Note that we take azimuthal coordinates in
$[-\pi,\pi]$ instead of the more conventional $[0,2\pi]$. This is done to keep the sequence of
$\Phi_{0,j}$ in strictly descending order.

\begin{table*}
  \begin{center}
    \begin{tabular}{c|cccccccc|}
      $j$   &   1      &  2  &  3  &   4  &   5  &   6  &   7  &  8  \\ \hline
      $b_j$ [$\upmu$G] & 0.1 & 3.0 & -0.9 & -0.8 & -2.0 & -4.2 & 0.0 & 2.7 \\
      $r_{-x,j}$ & 5.1 & 6.3 & 7.1 & 8.3 & 9.8 & 11.4 & 12.7 & 15.5 \\
      $f_j$ & 0.130 & 0.165 & 0.094 & 0.122 & 0.130 & 0.118 & 0.084 & 0.156 \\
      $\Phi_{1,j} / \pi $ &
      0.969 & 0.638 & 0.451 & 0.207 & -0.053 & -0.289 & -0.458 & -0.770 \\
      \hline
    \end{tabular}
    \caption{\label{tab:widths}
      Width parameters for the eight spiral regions. Rows 1 to 3 reproduced from JF12, row 4 from
      this work.
    }
  \end{center}
\end{table*}

\begin{figure}
  \begin{center}
    \includegraphics[width=0.9\columnwidth]{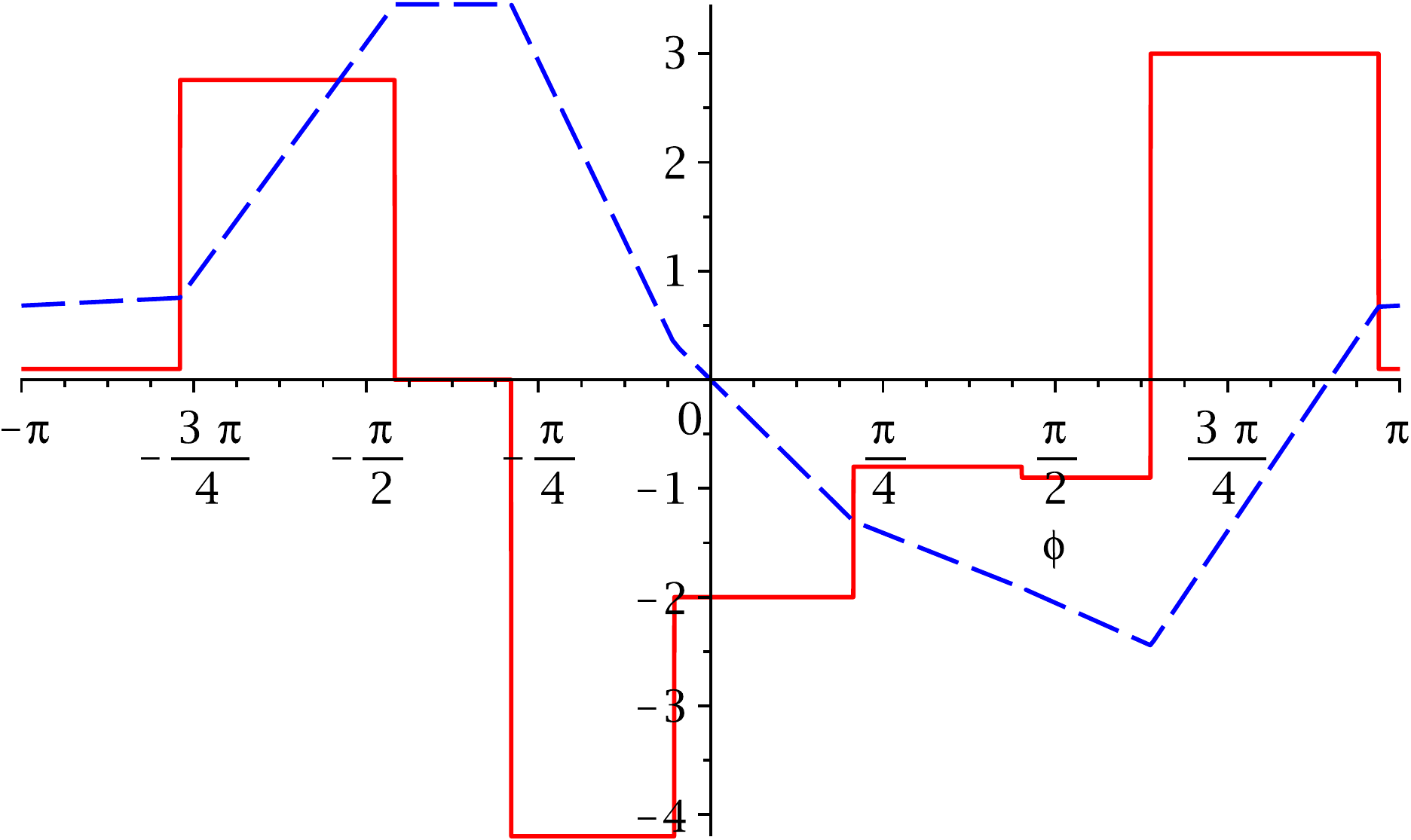}
    \caption{\label{fig:br-int}
      Step function $B_{r_1}(\vp)$ (red, solid) and its integral $H(\vp)$ (blue, dashed) as used
      in Equation~(\ref{eq:bp_cont}) evaluated at $r=r_1$, with lower interval bound $\vp_0=0$.
      A different bound $\vp_0^{\prime}$ would shift the blue curve vertically until
      $B_{r_1}(\vp_0^{\prime})=0$ is satisfied.
    }
  \end{center}
\end{figure}

The explicit magnetic field components at an arbitrary position $(r,\vp)$ within the spiral disk
can be obtained by first mapping the point along its field line back to the inner rim at
\begin{equation}
  (r_1, \vp_1) = \left( r_1, \vp - \frac{\ln (r/r_1)}{\tan i} \right) ,
\end{equation}
looking up the spiral region $j=j(\vp_1)$ that $\vp_1$ is situated in, and setting the field to
\begin{eqnarray}
  \label{eq:base_spiral-comps}
  [B_r, B_{\vp}] &=& \frac{b_j \, r_1}{r} [ \sin i, \cos i ] \\
  &=& B_{r_1} \left( \vp - \frac{\ln (r/r_1)}{\tan i} \right)
  \frac{r_1}{r} \, [ \sin i, \cos i ] \nonumber
\end{eqnarray}
where $B_{r_1}(\vp)$ is the periodic step function that maps
\mbox{$\vp \in [\Phi_{1,j-1}, \Phi_{1,j}]$} to $b_j$, see Figure~\ref{fig:br-int}. For heights
\mbox{$z>0$} \mbox{($z<0$)} above (below) the Galactic plane, an additional factor,
\begin{equation}
  \label{eq:Lminus1}
  \Lambda(z) \equiv 1-L(z,h,w) \equiv \left[1+\exp\left(\frac{|z|-h}{w/2}\right)\right]^{-1} ,
\end{equation}
with parameters \mbox{$[h,w]=[0.4, 0.27]$~kpc} is added to the right-hand side of
Equation~(\ref{eq:base_spiral-comps}). Since this factor does not depend on $(r,\vp)$, and is
therefore immaterial to the question of magnetic flux conservation, it will be neglected in the
following, thus restricting our ensuing considerations of the spiral disk to the Galactic ($z=0$)
plane.

The above formulation has the clear advantage that the solenoidal correction that will be
described in the next section can easily be applied to an existing implementation of the JF12
field that should have $B_{r_1}(\vp) = \|\vec{B}\|_{r=r_1}$ readily accessible. \\

\subsection{Clearing Divergences at the Spiral Boundaries}
\label{sect:fix_spirals}

The general idea behind our proposed method to make the spiral field fully divergence-free is to
first define a new parameter $\delta>0$ and to use the unmodified spiral field only within the
central part of the disk at \mbox{$r \in [r_1+\delta, r_2-\delta]$}, while the radial intervals
\mbox{$[r_1, r_1+\delta]$} and \mbox{$[r_2-\delta, r_2]$} form circular annular regions of width
$\delta$ at both boundaries, in which incoming and outgoing flux is smoothly redistributed.
For simplicity of the argument, we take the values of $\delta$ at the inner and outer boundary to
be the same, but still note that they could easily be chosen differently for a given application.
For the outer rim $r_2$, an alternative treatment not involving a transition region will be
described toward the end of this section.

\begin{figure}
  \includegraphics[width=\columnwidth]{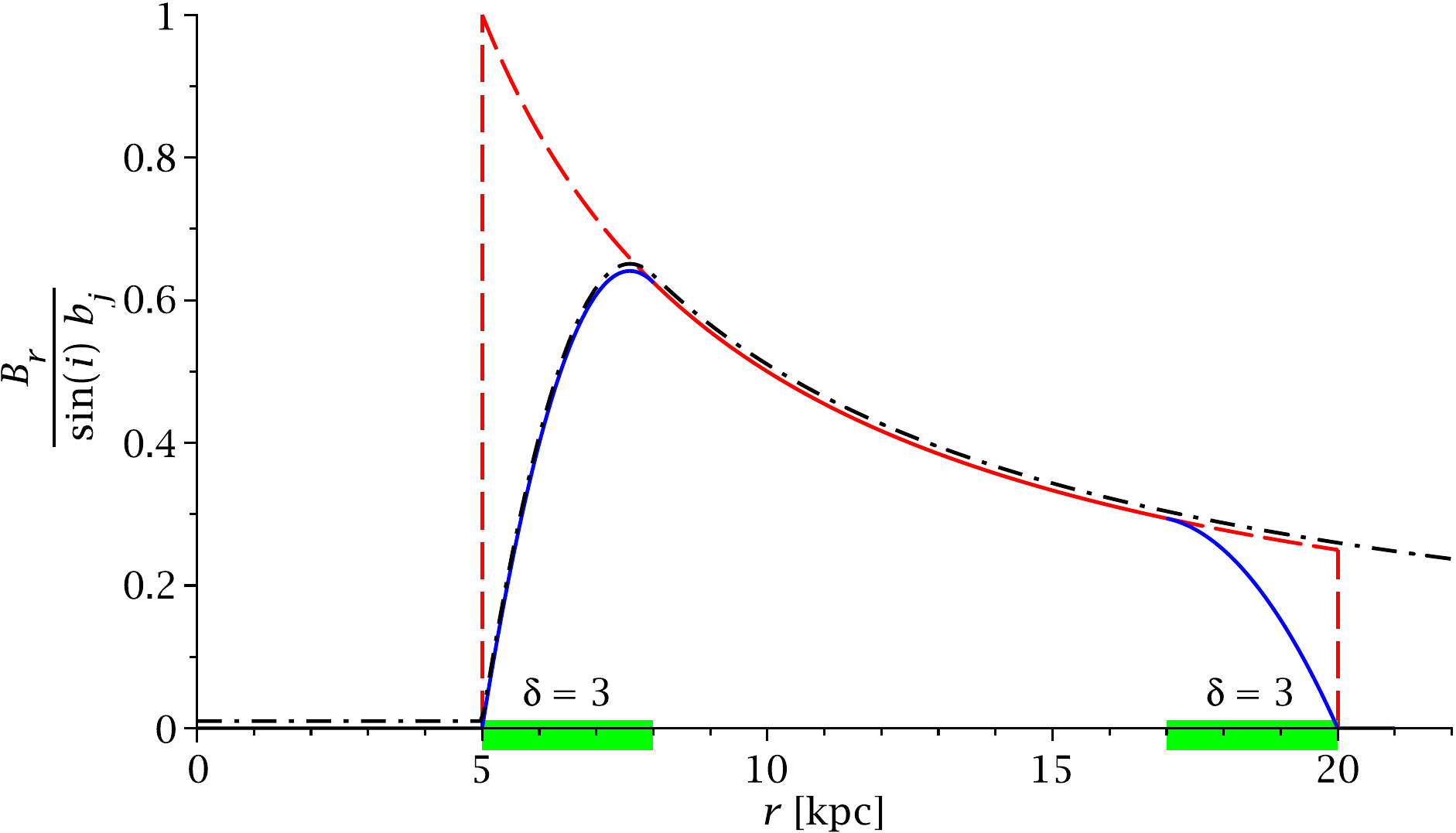}
  \caption{\label{fig:p_delta}
    JF12's normalized $B_r$ profile $r_1/r$ (red dashed and dotted) and polynomial inserts
    $p_{\delta}(r)$ (blue) within the two transition regions (indicated by green bars) of width
    $\delta=3$~kpc. The new total compound profile (employing azimuthal flux redistribution at
    both the inner and outer edges) is composed of the respective contributions drawn in solid
    linestyle (red and blue). As an alternative, the spiral disk could be extended to infinity
    via \mbox{$r_2 \rightarrow \infty$} (black, dash-dotted). \\
  }
\end{figure}

Spiral field lines traversing radii $r_1+\delta$ and $r_2-\delta$ do so at inclination $i$, and
are to be smoothly continued into the respective transition regions. In a first step, the factor
$r_1/r$ in Equation~(\ref{eq:base_spiral-comps}) is replaced by a polynomial $p_\delta(r)$ inside
the transition regions, leading to
\begin{eqnarray}
  \label{eq:B_sp}
  [ \bar{B}_r, \bar{B}_{\vp} ] &=&
  B_{r_1} \left( \vp - \frac{\ln (r/r_1)}{\tan i} \right) p_{\delta}(r)
  \, [ \sin i, \cos i ] .
\end{eqnarray}
From here onwards, a symbol with a bar denotes quantities introduced in addition to JF12, while
those without a bar are the original ones from that paper. Here, $p_{\delta}(r)$ is a second-order
polynomial whose coefficients are fixed by requiring $\bar{B}_r$ to be differentiable at the
limiting radius \mbox{$\rbet = r_1+\delta$} (\mbox{$\rbet = r_2-\delta$}), separating the
intermediate, unmodified region from the inner (outer) transition region, and to vanish at
$r=r_1$ ($r=r_2$), where the entire spiral disk ends. These requirements result in the explicit
expression
\begin{equation}
  \label{eq:p_delta}
  p_{\delta}(r) = \frac{r_1}{\rbet} \left[
    2 -\frac{r}{\rbet}
    + \left( \frac{\ralp}{\rbet}-2 \right) \left(\frac{r-\rbet}{\ralp-\rbet}\right)^2
    \right]
\end{equation}
at the inner ($\ralp=r_1$, $\rbet=r_1+\delta$) and outer ($\ralp=r_2$, $\rbet=r_2-\delta$) rim of
the spiral disk. Figure~\ref{fig:p_delta} serves to illustrate the situation.

Up until now, we have merely modified the peripheral regions of the spiral disk in a way that lets
its field strength smoothly tend to zero while avoiding kinking field lines. In order to determine
the additional $\vp$ component $B_{\vp,{\rm add}}$ that ensures magnetic solenoidality within the
transition regions, we use
\begin{equation}
  0 = \nabla \cdot \bar{\vec{B}} = \frac{1}{r} \left(
  \pdv{(r \, \bar{B}_r)}{r} + \pdv{(\bar{B}_{\vp} + \bar{B}_{\vp,{\rm add}})}{\vp} \right)
\end{equation}
and Equation~(\ref{eq:B_sp}) to get
\begin{equation}
  \label{eq:bp_cont}
  \bar{B}_{\vp,{\rm add}} = -\underbrace{\pdv{[r \, p_{\delta}(r)]}{r}}_{\equiv q_{\delta}(r)} \
  H \left(\vp - \frac{\ln (r/r_1)}{\tan i} \right) \sin i
\end{equation}
with
\begin{equation}
  \label{eq:def_intH}
  H(\vp) = \int_{\vp_0}^{\vp} B_{r_1} (\vp\p) \dd \vp\p .
\end{equation}
The first factor of Equation~(\ref{eq:bp_cont}) can be evaluated straightforwardly from
Equation~(\ref{eq:p_delta}) as
\begin{equation}
  \label{eq:q_delta}
  q_{\delta}(r) = \frac{r_1}{\rbet} \left[ 2 -2 \frac{r}{\rbet}
    + \left( \frac{\ralp}{\rbet}-2 \right)
    \frac{3 r^2-4 \rbet r+\rbet^2}{(\ralp-\rbet)^2} \right]
\end{equation}
and the integral $H(\vp)$ yields a piecewise linear, $2\pi$-periodic function of $\vp$. The lower
interval bound $\vp_0$ may be interpreted as the azimuthal direction at which an impenetrable wall
with inclination $i$, separating magnetic flux being redirected into clockwise and
counterclockwise directions, intersects $r=r_1$. Its value may be chosen freely, one possibility
being a choice that minimizes the maximum or average value of additional azimuthal magnetic flux
or energy that is brought into the system. For simplicity, Figure~\ref{fig:br-int} uses the ad~hoc
value $\vp_0=0$ that apparently results in a rather balanced distribution.
In total, the transition zone field is to be set as
\begin{equation}
  \bar{\vec{B}}_{\rm tr} = \bar{B}_r \, \vec{e}_r
  + \left[ \bar{B}_{\vp} + \bar{B}_{\vp, {\rm add}} \right] \vec{e}_{\vp}
\end{equation}
with components given by Equations~(\ref{eq:B_sp}) and (\ref{eq:bp_cont}).

Identifying the most appropriate choice for the transition thickness $\delta$ is not
straightforward. A small value will leave most of the disk field unchanged, which could be a
desirable strategy in order to minimize interference with the delicate agreement with observational
data. On the other hand, the combined azimuthal flux of field lines being tightly packed into two
very thin transition zones may then become unreasonably large. For the intermediate value of
$\delta=3.0$~kpc, the added azimuthal flux from $\bar{B}_{\vp, {\rm add}}$ is comparable to
the reduction in spiral flux which arises due to the magnitude of $p_{\delta}(r)$ being
considerably smaller than $r_1/r$ in the region of interest, as can clearly be seen in
Figure~\ref{fig:p_delta}.

Since the presented method of flux redistribution inside the disk is of course not the only way
to ensure a divergence-free field, one may ask at this point how it compares to other strategies.
For instance, one could also divert field lines away from the Galactic plane and into the halo,
similarly to how \citet{Ferriere_Terral:2014} avoid infinite field strengths at their model's
polar axis. Our reasons for considering the disk in isolation is that it allows us to stay
conceptually closer to the JF12 model, and in particular to take advantage of the total balance of
incoming and outgoing flux expressed in Equation~(\ref{eq:normflux}).

While some form of flux rearrangement is inevitable at the inner rim, yet a different option
could be exploited at the outer rim by simply moving its position from \mbox{$r_2 = 20$~kpc} to
infinity (or, from a more practical point of view, beyond the specific boundaries of the region
under consideration), such that the field strength continues to decay as $1/r$ indefinitely.
Since with this radial profile, the field strength at, say, \mbox{$r=25$~kpc} would still amount
to 20\% of its reference value at $r_1$, it is clear that either option would represent a marked
deviation from the original JF12 disk field. A reassessment via fitting to observables, ideally
including both $\delta$ and $r_2$ as yet two more free parameters will therefore in any case be
mandatory, and in this sense, the two possibilities outlined above represent the limiting cases in
$(\delta, r_2)$ parameter space. This task, however, is beyond the scope of this present work,
which merely seeks to present and discuss a subset of physically admissible options.

\begin{figure*}
  \centering
  \includegraphics[width=\textwidth]{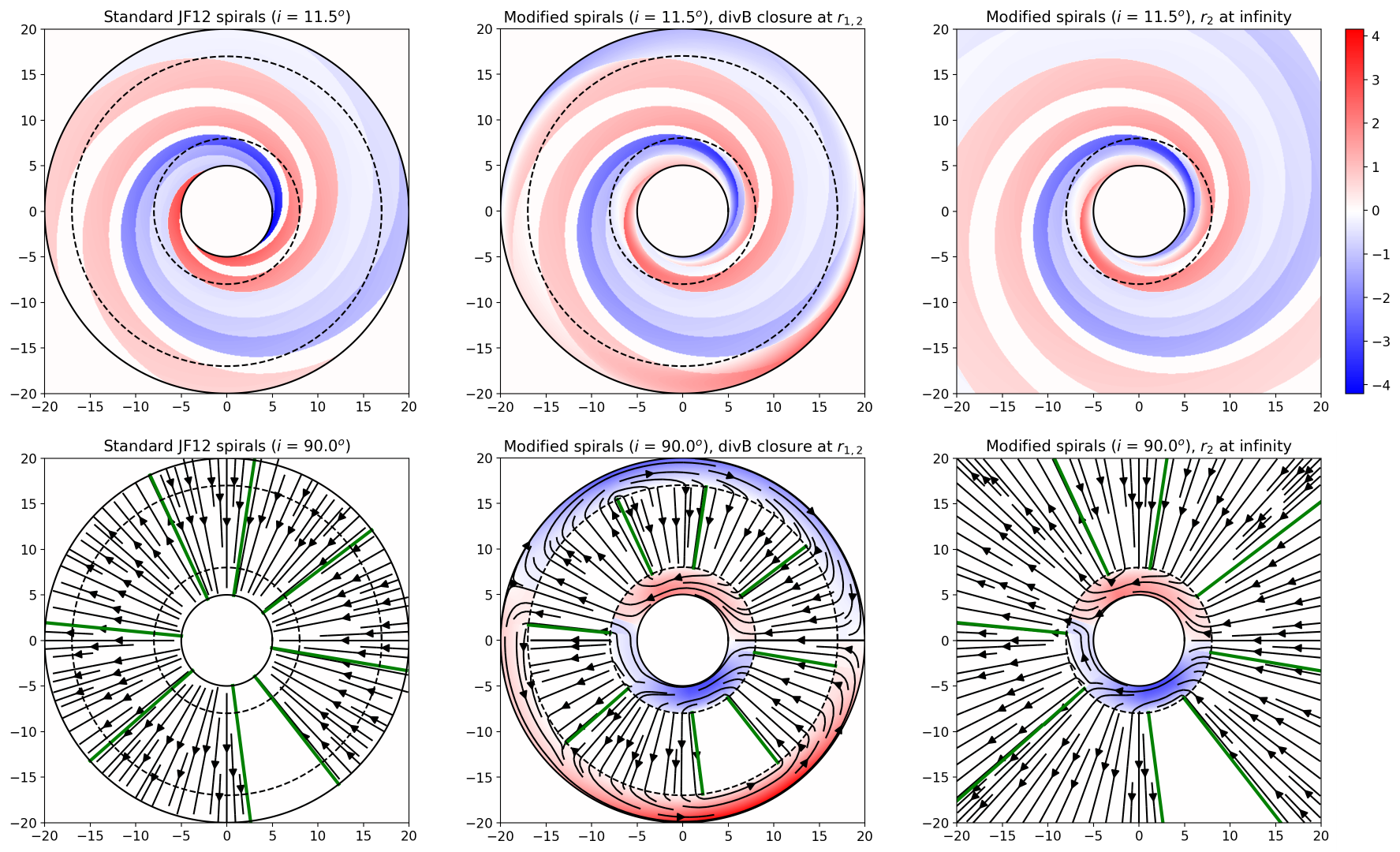}
  \caption{\label{fig:compare_diskfield}
    Contour plots of magnetic field strength multiplied with the sign of $B_\vp$, such that
    negative (positive) values indicate clockwise (counterclockwise) field orientation. The top
    row uses the actual inclination angle of $i=11.5^{\circ}$, while the bottom row shows segments
    of field lines for the hypothetical case of $90^{\circ}$ inclination, illustrating the concept
    of flux redistribution in the inner and outer transition regions, whose width is uniformly
    chosen as $\delta=3$~kpc. Green radial lines visually separate the eight ``spiral'' regions,
    across which the field orientation may change from inward- to outward-directed.
    Left: the original JF12 configuration, essentially reproducing the upper left plot of Figure~7
    from that paper.
    Middle: the modified, and thus fully solenoidal, field for a transition width of
    $\delta=3$~kpc (bounded by dashed lines).
    Right: an alternative treatment for the outer rim, whose position $r_2$ moves from 20~kpc to
    infinity, allowing field lines (of vanishing field strength) to ``close'' there.
    Note the straight field line at $\vp=\varphi_0=0 \Leftrightarrow y=0, x>0$ separating left-
    and right-going flux. The fact that field lines seem to have endpoints is an artifact of the
    employed plotting procedure. Note also that the color bar is not relevant for the bottom row
    of plots, whose maximum absolute field strength is larger than that of the other two cases by
    a factor of about three. \\
  }
\end{figure*}

While we do acknowledge that an indefinite $1/r$ decay is a simple and widely accepted possibility
in the community, one should be very aware of the conceptual and practical implications of a
galactic disk whose magnetic field is truly unbounded in spatial extent. Specifically, the total
energy
\begin{equation}
  W_{B, \textrm{disk}} \propto \int_{-\infty}^\infty \int_{r_1}^{r_2}
  \left(\frac{\Lambda(z)}{r}\right)^2 r \dd r \dd z = C \ln \left(\frac{r_2}{r_1} \right)
\end{equation}
contained in the magnetic field of such a disk (with \mbox{$C\approx 0.557$} arising from vertical
integration) obviously diverges as \mbox{$r_2 \rightarrow \infty$}, leading to what could be
called a ``magnetic Olbers' paradox.'' Even if this limit may not actually be realized in most
practical applications, it still seems conceivable that the excess of energy thus implied may have
a distorting effect on, for instance, cosmological simulations involving large volumes populated
with galactic disks, or line-of-sight integrations connecting the observer to distant galaxies.
Figure~\ref{fig:compare_diskfield} compares the old and new spiral field structure for both
strategies, and also uses a second row of plots to illustrate the general idea of flux being
redistributed.

These plots seem to suggest that the unmodified JF12 spiral field is recovered in
the limit $\delta \rightarrow 0$. While this is indeed the case within the open annulus
\mbox{$r_1 < r < r_2$}, the additional azimuthal flux would then accumulate to form a singular,
infinitely strong flux ring in the transition region of zero width, which would be just as
unphysical (although for a different reason) as cutting all field lines at the radial boundaries.

We note that, as can be seen by carefully inspecting the lower middle plot of 
Figure~\ref{fig:compare_diskfield}, field lines may kink when crossing the boundary between spiral
regions. This is unavoidable near $r=\rbet$ due to the discontinuous transitions between these
regions that are an inherent feature of the JF12 model. Further into the transition regions, this
could in principal be avoided by replacing the piecewise constant integrand in
Equation~(\ref{eq:bp_cont}) by a smoothed version of itself in a way that increases the smoothing
length from zero at $\rbet$ to a finite value toward $\ralp$. However, presenting and discussing
appropriate formulas to this end is beyond the scope of this paper as well. \\

\section{Improvements for the X-type Field Component}
\label{sec:x-type_field}

In this section, we turn our attention away from the disk and toward the poloidal component,
first noting another divergence-related problem with the latter, which is fortunately much easier
to solve. The remainder of the section deals with the problem of kinking field lines and
discusses two appropriate mitigation strategies. \\

\subsection{Explicit Formulas}

As with the spiral field, we also briefly list explicit formulas for the components $[B_r, B_z]$
of the poloidal X-field, both for later reference and because not all of them have been provided
explicitly in the literature. The X-field is characterized by straight field lines, whose elevation
angle \mbox{$\Tet \equiv \mathrm{arctan}(B_z/B_r)$} varies as follows.
A field line intersecting the midplane $z=0$ at radius $\rp$ has
\begin{equation}
  \label{eq:jf12_tet}
  \tan \Tet = \tan \Tetn \times \left\{ \begin{array}{ccl}
    \rxc / \rp &:& \rp  <  \rxc \\
    1          &:& \rp \ge \rxc
  \end{array} \right.
\end{equation}
with global constants $\rxc$ and $\Tetn$, and $B_r$ changes sign at the midplane, such that
$B_r / z > 0$. Note that, according to this equation, it is $\cot \Tet$, rather than the
inclination $\Tet$ itself, that depends linearly on $\rp$ in the ``linear'' region.

The easiest way to obtain cylindrical components $[B_r, B_z]$ at given $(r,z)$ is to first compute
$\tilde{r}_{\rm p}\equiv r - |z| \cot \Tetn$ and then define
\begin{equation}
  \rp \equiv \left\{ \begin{array}{ccc}
    \tilde{r}_{\rm p} &:& r > \tilde{r}_{\rm p} \\
    r \, \rxc/(\rxc+|z| \cot \Tetn) &:& r \le \tilde{r}_{\rm p}
  \end{array} \right.
\end{equation}
and further
\begin{equation}
  \label{eq:Bcomps_X}
  [B_r, B_z] \equiv \frac{b_{\rm X}(\rp)}{(r/\rp)^d}
  \left[ \mathrm{sgn}(z) \cos \Tet, \sin \Tet \right]
\end{equation}
with an absolute field strength
\begin{equation}
  \label{eq:Bat0_jf12}
  b_{\rm X}(\rp) = B_{\rm X} \exp (-\rp/\rx)
\end{equation}
at $z=0$, and an exponent
\begin{equation}
  \label{eq:def_d-exponent}
  d\equiv \left\{ \begin{array}{ccl}
    2 &:& \rp  <  \rxc \\
    1 &:& \rp \ge \rxc
  \end{array} \right.
\end{equation}
that reflects the different scaling, which the solenoidality condition (\ref{eq:divB0}) enforces
in the two regions. \\

\subsection{Solenoidality Near the Origin and at Large Distances}

The X-type field of JF12 omits a spherical region of radius 1~kpc around the origin, in which the
field is set to zero. We note, however, that despite the seemingly divergent scaling of
Equation~(\ref{eq:Bcomps_X}) as $r \rightarrow 0$, the field components remain perfectly
well-defined also on the $z$-axis, where they smoothly converge to 
\begin{equation}
  \lim_{r \rightarrow 0} \ [B_r, B_z] = \left[ 0, \, B_{\rm X}
    \left(1+ \frac{|z|}{\rxc \tan \Tetn}\right)^{-2} \right]
\end{equation}
with the field strength attaining its global, finite maximum value of $B_{\rm X}$ right at the
origin. This can be seen by noting that, according to Equation~(9) in JF12, $\rp/r$ is actually
independent of $r$ in the ``linear'' region and may be straightforwardly evaluated at any $z$,
including $z=0$.
For this reason, we assume in the following that the field (\ref{eq:Bcomps_X}) is being continued
also inside the previously excluded central region within 1~kpc of the origin. This is justified
not only by simplicity (and, above all, as the canonical means to restore the otherwise
violated solenoidality at the surface of the ``exclusion sphere''), but also by to the fact that,
as stated in JF12, the employed RM data does not permit to constrain the central part of the field
in a useful way.

Finally, we note that because the JF12 X-field, just like its spiral disk field, is set to zero
outside a cylinder of 20~kpc radius around the $r=0$ axis, the question again arises as to how
solenoidality should best be restored at this outer cylindrical rim surface. One straightforward
possibility is to depart from the original prescription by \textit{not} cutting the field at a
finite cylindrical radius and instead letting its field strength decay as $1/r$ (or $1/r^2$,
depending on direction) instead. The second alternative would again consist of a redistribution of
field lines that would then likely assume a dipole-like structure, as flux needs to be transported
from the Northern to the Southern Galactic half-space. Since there are currently no observational
indications for a departure from the X-shape structure at large distances from the Galactic
center, we refrain from making explicit suggestions for such a formal closure at this point.
We merely note that, as with the spiral field, some form of field line closure is clearly
desirable -- if not mandatory -- for fundamental reasons, and that the discrimination between
different methods and their respective parameters should again be inferred based on observational
constraints. \\

\subsection{Motivating the Need for a Kink-free X-field}

The inclusion of the X-type field into JF12 was motivated by corresponding radio observations of
edge-on galaxies \citep[e.g.][]{Beck:2009}, and the simplest way to model this feature is to
employ field lines which are straight on either side of the Galactic plane, where they meet to
form wedge-shaped kinks.
While this may be a very reasonable approximation for many applications, it does harbor problems
for others. For instance, the associated current sheet of infinite strength at the $z=0$ plane is
unphysical, and will thus tend to generate equally unphysical forces in MHD simulations. To see
this, we may approximate the kink as the limiting case of a smooth, X-shaped, and dimensionless
poloidal field
\begin{equation}
  \vec{B}^{\star} = \arctan(z/\eta) \, \vec{e}_r + \vec{e}_z ,
\end{equation}
which has a finite radius of curvature $\eta$ at $z=0$, as well as an -- for this purpose
irrelevant -- asymptotic inclination angle of $\arctan(2/\pi) \approx 32^{\circ}$. At \mbox{$z=0$},
the respective dimensionless expressions for the associated densities of electric current and
Lorentz force are then found from
\begin{eqnarray}
  \vec{J}^{\star}|_{z=0} \ &=& \left. (\nabla \times \vec{B}^{\star}) \right|_{z=0}
  = \left. \frac{\eta}{\eta^2+z^2} \right|_{z=0} \hspace*{-2mm} \vec{e}_\vp
  = \frac{1}{\eta} \, \vec{e}_\vp \\
  \vec{F}_{\rm L}|_{z=0} \ &=& \left. (\vec{J}^{\star} \times \vec{B}^{\star}) \right|_{z=0}
  = \frac{1}{\eta} \, \vec{e}_r ,
\end{eqnarray}
both of which diverge as $\eta \rightarrow 0$. Note that this line of reasoning is not affected
by the nonzero value of $\nabla \cdot \vec{B}^\star$.

Another instance in which smooth field lines are preferred over kinking ones is the numerical
tracing of charged particles, where the necessarily finite step size makes it difficult to
accommodate rapid or even discontinuous changes in field strengths along the trajectory of a
particle that would otherwise simply follow its original field line.

With this motivation in mind, we now proceed to present two modifications that keep the radius of
curvature finite within a planar region around the midplane, while the field outside this region
is largely left unchanged, with field lines smoothly traversing the boundaries between those
regions. This ensures that the desired result is obtained while again keeping the unavoidable
interference with JF12's fine-tuned set of parameters at a minimum. \\

\subsection{Method I: Parabolic Replacement Near the Disk}
\label{sec:parabolic}

\subsubsection{General Idea and Formulas}

We first consider the more general case of a largely arbitrary source field $\vec{B}$, which we
merely require to obey symmetry relations $B_r(r,-z)=-B_r(r,z)$ and $B_z(r,-z)=B_z(r,z)$, and only
later specialize to the JF12 X-field. The goal is to leave $\vec{B}$ unchanged outside a freely
chosen reference height $|z| \ge \zs>0$ (thereby ensuring that the original $\vec{B}$ is fully
recovered in the limit \mbox{$\zs \rightarrow 0$}), but create a replacement field
$\bar{\vec{B}}$ inside \mbox{$|z|<\zs$} whose field lines are given by parabolas
\begin{equation}
  \rF(\rs, z) \equiv a(\rs) + b(\rs) z^2
\end{equation}
which are parameterized by the radius $\rs$ at which the respective field line passes $|z|=\zs$,
smoothly connecting to its outer counterpart. This parameterization is analogous to the one using
$\rp$ (except for the finite, rather than zero, reference height), and both are in fact related
via
\begin{equation}
  \label{eq:para_link}
  (\rs - \rp) \tan \Tet = \zs .
\end{equation}
The coefficient functions $a(\rs)$ and $b(\rs)$ are fixed by requiring that field lines be
continuous and differentiable at height $\zs$ via
\begin{eqnarray}
  \rF(\rs, \zs) & \ = \ & \rs \\
  \left. \pdv{\rF(\rs,z)}{z} \right|_{\zs} &=& \left. \frac{B_r}{B_z} \right|_{(\rs,\zs)}
\end{eqnarray}
yielding
\begin{equation}
  \label{eq:fieldline}
  \rF(\rs, z) = \rs - \frac{1}{2}
  \left(\zs-\frac{z^2}{\zs}\right) \frac{B_r(\rs,\zs)}{B_z(\rs,\zs)} .
\end{equation}
We then once more use the definition of field lines (this time for $|z|<\zs$) to obtain
\begin{equation}
  \label{eq:ratio}
  \frac{\bar{B}_r\Big(\rF(\rs,z), z\Big)}{\bar{B}_z\Big(\rF(\rs,z), z\Big)}
  = \frac{\partial \rF(\rs, z)}{\partial z} = 
  \left(\frac{z}{\zs}\right) \frac{B_r(\rs,\zs)}{B_z(\rs,\zs)}
\end{equation}
by differentiating our newly found Equation~(\ref{eq:fieldline}). We see that indeed,
$\bar{B}_r \rightarrow 0$ as $|z| \rightarrow 0$, and also that $\bar{B}_r$'s change of sign at
the midplane is maintained.

Next, we exploit the divergence constraint by considering the conservation of magnetic flux
\begin{equation}
  \label{eq:equal_flux}
  2\pi \, r \, \dd r \ \bar{B}_z(r,z) =
  2\pi \, \rs \, \dd \rs \ B_z(\rs,\zs)
\end{equation}
from an arbitrary height $z < \zs$ to $z=\zs$ through a circular, disk-parallel annulus of
infinitesimal radial width $\dd r$ along a field line passing through a given position $(r,z)$.
Here, $\rs$ is the parameter of the parabola passing through $(r,z)$, and is therefore to be
obtained from the condition $r=\rF(z)$ using Equation~(\ref{eq:fieldline}).
At height $z$, the radial width of the annulus bounded by adjacent parabolic field lines
$\rs$ and $\rs+\dd \rs$ is
\begin{equation}
  \dd r = \rF(\rs + \dd \rs, z) - \rF(\rs, z)
  = \frac{\partial \rF(\rs,z)}{\partial \rs} \dd \rs
\end{equation}
when neglecting terms of order ${\cal O}(\dd \rs^2)$. Therefore, Equation~(\ref{eq:equal_flux})
implies
\begin{eqnarray}
  \label{eq:Bz_gen}
  F(r,z,\rs) &\equiv& \frac{\bar{B}_z(r,z)}{B_z(\rs,\zs)} = \frac{\rs/r}{ \displaystyle
    \left( \frac{\partial \rF(\rs,z)}{\partial \rs} \right)} \\
  \nonumber &=& \frac{ \displaystyle 1 + \frac{1}{2 r}
    \left(\zs-\frac{z^2}{\zs}\right) \frac{B_r(\rs,\zs)}{B_z(\rs,\zs)}}{ \displaystyle
    1 - \frac{1}{2}
     \left(\zs-\frac{z^2}{\zs}\right) \frac{\partial}{\partial \rs}
     \left[\frac{B_r(\rs,\zs)}{B_z(\rs,\zs)}\right] }
\end{eqnarray}
and, together with Equation~(\ref{eq:ratio}),
\begin{equation}
  \label{eq:Bcomp_gen}
  \begin{split}
    \bar{\vec{B}}(r,z) =& \left[ \left(\frac{z}{\zs}\right)
      B_r(\rs,\zs) \, \vec{e}_r + B_z(\rs,\zs) \, \vec{e}_z \right] \\
    & \times F(r,z,\rs) .
  \end{split}
\end{equation}
Further evaluation of this equation is precluded by the fact that the implicit
Equation~(\ref{eq:fieldline}) cannot be solved for $\rs$ in this general form. \\

\subsubsection{Application to JF12}

Using Equation~(\ref{eq:jf12_tet}) for the case of JF12, the field line equation
(\ref{eq:fieldline}) becomes
\begin{equation}
  \label{eq:fieldline_jf12}
  \begin{split}
    \rF(\rs, z) =&\ \rs -
    \frac{\zs}{2 \tan \Tetn} \left(1-\frac{z^2}{\zs^2}\right) \\
    & \times \left\{ \begin{array}{ccl}
      \rs / \rsc &:& \rs  <  \rsc \\
      1          &:& \rs \ge \rsc
    \end{array} \right.
  \end{split}
\end{equation}
when expressed in terms of $\rs$. Here,
\begin{equation}
  \rsc\equiv \rxc + \zs/\tan \Tetn
\end{equation}
is the radius at which the ``critical'' straight field line, defined as the one separating both
regions and crossing $z=0$ at radius $\rxc$, intersects the $z=\zs$ plane. We can see from
Equation~(\ref{eq:fieldline_jf12}) that in the outer region ($r \ge \rsc$), the parabolas are
identical except for a translation in $r$, while in the inner region, they are additionally
compressed in the $r$ direction, becoming straight and vertical at the \mbox{$r=0=\rs$} axis.

To construct the new field $\bar{\vec{B}}$ at position $(r,z)$ within \mbox{$|z| \le \zs$}, we
first need to find the parameter $\rs$ of the corresponding parabola. Assuming $\rs \ge \rsc$ in
Equation~(\ref{eq:fieldline_jf12}), the condition $r=\rF(\rs,z)$ may be trivially solved for
$\rs$, giving
\begin{equation}
  \label{eq:solved_r0_out}
  \rs =  r + \frac{\rxc}{\beta_0} \left(1-\frac{z^2}{\zs^2}\right)
\end{equation}
with $\beta_0 \equiv 2 (\rxc \tan\Tetn)/\zs$ as a constant.
If the assumption $\rs \ge \rsc$ turns out to be correct for the point in question,
Equation~(\ref{eq:solved_r0_out}) gives the desired $\rs$, or else Equation~(\ref{eq:fieldline_jf12})
points us to
\begin{equation}
  \label{eq:solved_r0_in}
  \rs = r \left[ 1- \frac{1}{2 + \beta_0}
    \left(1-\frac{z^2}{\zs^2}\right) \right]^{-1}
\end{equation}
for the ``inside'' case. Using relation (\ref{eq:para_link}) between $\rp$ and $\rs$, we have for
the inner region
\begin{equation}
  \frac{1}{\tan\Tet} = \frac{1}{\tan\Tetn} \, \frac{\rp}{\rxc}
  =  \frac{1}{\rxc \tan\Tetn} \left( \rs - \frac{\zs}{\tan \Tet}\right) ,
\end{equation}
which may be solved to yield
\begin{equation}
  \frac{1}{\tan \Tet} = \frac{\rs}{\rxc \tan \Tetn + \zs} ,
\end{equation}
and further
\begin{equation}
  \begin{split}
    \frac{\partial}{\partial \rs} \left[\frac{B_r(\rs,\zs)}{B_z(\rs,\zs)} \right] 
    =&\ \frac{\partial}{\partial \rs}  \left[\frac{1}{\tan\Tet|_{(\rs,\zs)}}\right] \\
    =&\ \frac{1}{\rxc \tan \Tetn + \zs} .
  \end{split}
\end{equation}
In the outer region, where \mbox{$\Tet=\Tetn$} is a constant, this derivative vanishes.
Finally, we are ready to fully evaluate Equation~(\ref{eq:Bcomp_gen}) and determine $F$ as
\begin{equation}
  \label{eq:Bcomp_jf12_in_out}
  F(r,z) = \left\{
    \begin{array}{ccl}
      \displaystyle \left[1-\frac{1}{2+\beta_0} \left(1-\frac{z^2}{\zs^2} \right) \right]^{-2}
      &:& \mbox{inside} \\ && \\
      \displaystyle 1 + \frac{1}{\beta_0} \left(\frac{\rxc}{r}\right)
      \left(1-\frac{z^2}{\zs^2}\right) &:& \mbox{outside}
    \end{array} \right.
\end{equation}
in the two regions. The third argument $\rs$ in $F$ has now been suppressed because $\rs=\rs(r,z)$
was inserted from Equation~(\ref{eq:fieldline_jf12}).

\begin{figure*}[h]
  \centering
  \includegraphics[width=0.98\textwidth]{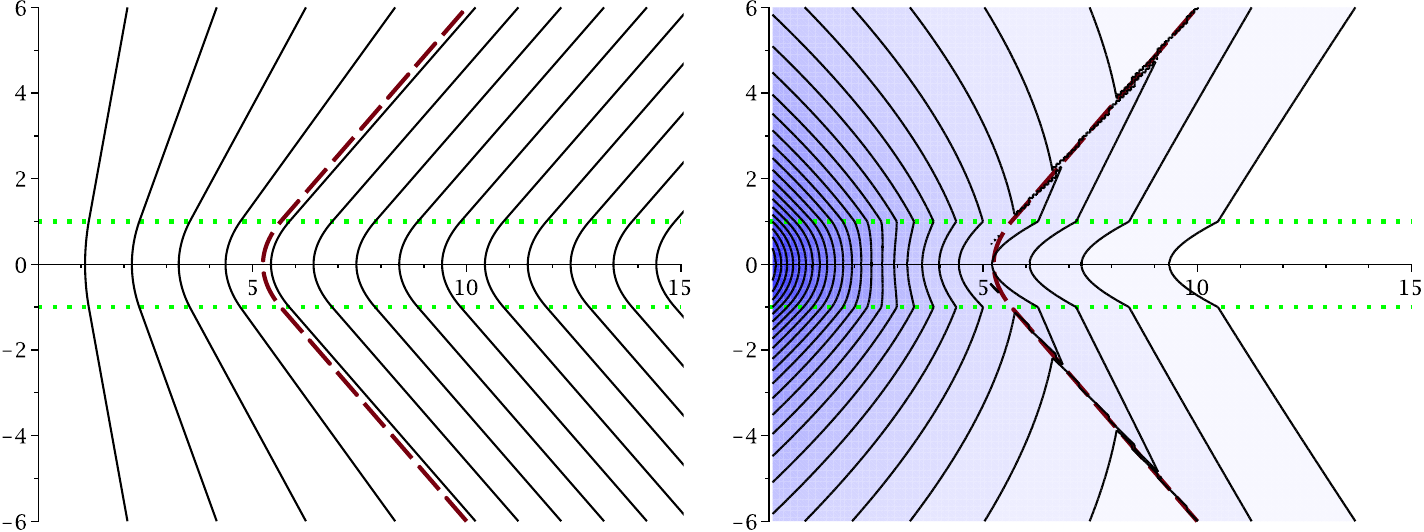}
  \caption{\label{fig:xfield}
    Field lines (left) and contours of $\|\vec{B}\|$ (right) in the poloidal ($r,z$) plane.
    The formerly wedge-shaped field (see Figure~5 in JF12) is smoothened within the
    \mbox{$|z| \le \zs=1$} region (bounded by the green dotted lines). Note that the jump in field
    strength at the inner-outer separator field line (brown dashed line) is induced by the
    different scalings ($\propto r^{-1}$ vs.\ $r^{-2}$) of both regions according to
    Equations~(\ref{eq:Bcomps_X}) and (\ref{eq:def_d-exponent}), and is therefore already present
    in the original JF12 field.
  }
\end{figure*}

In summary, the procedure to evaluate the improved field at arbitrary $(r,z)$ for a global choice
of $\zs$ is as follows:
\begin{enumerate}
\item Discriminate between ``inner'' and ``outer'' region as before, but replacing the criterion
  $\rp < \rxc$ by $\rs < \rsc$ within $|z| < \zs$.
\item If $|z| < \zs$, compute $\rs$ using either Equation~(\ref{eq:solved_r0_out}) or
  (\ref{eq:solved_r0_in}), depending on whether $\rs \ge \rsc$ or not.
\item Compute the standard field at $(\rs,\zs)$, then the new field $\bar{\vec{B}}$ at $(r,z)$
  using Equations~(\ref{eq:Bcomp_gen}) and (\ref{eq:Bcomp_jf12_in_out}).
\end{enumerate}
Figure~\ref{fig:xfield} serves to illustrate the result thus obtained. \\

\subsection{Method II: Smoothing via Convolution}
\label{sec:smoothing}

A ``global'' alternative to the smoothing approach which was previously discussed is a convolution
of the JF12 X-field with a smooth kernel function $K \in C_0^{\infty}(\mathbb{R}^3)$, a so-called
``mollifier.'' The convolution of the Cartesian field components $B_c$, $c \in \{ x,y,z\}$ is
given as
\begin{equation}
  \begin{split}
    \label{eq:convdef}
    \tilde{B}_c(\mathbf{r}) &= (B_c * K)(\mathbf{r}) \\
    &= \int_{\text{supp}(K)}{B_c(\mathbf{r} - \mathbf{r\p}) \, K(\mathbf{r\p})
      \, \dd^3 r\p} .
  \end{split}
\end{equation}
This integral operation will always yield a smooth $C^{\infty}$ field if the initial field is
locally integrable, so this method is not restricted to the field configuration at hand.
Furthermore, it preserves the solenoidality of the initial field, which may be checked using the
identity
\begin{align}
  \pdv{}{x_k} (B_c * K) (\mathbf{r}) = \left( \pdv{B_c}{x_k} * K \right)(\mathbf{r})
\end{align}
that holds for any differentiable function $B_c$ within the compact support of $K$. We use the
standard mollifier
\begin{align}
  K(\mathbf{r}) = \left\{
  \begin{array}{lcl}
    \displaystyle \mathcal{N} \exp \left[
    \frac{1}{(\| \vec{r} \|/ w_{\text{X}})^2-1} \right] &:&
    \| \mathbf{r} \| < w_{\text{X}} \\
    0 &:& \| \vec{r}\| \geq w_{\text{X}}
  \end{array} \right.
\end{align}
where $w_{\textrm{X}}$ denotes the radius of the kernel's compact support, and $\mathcal{N}$
normalizes the function. The convolution averages the initial field inside a sphere of radius
$w_{\textrm{X}}$ with $K$ as a weight function. As it is not possible to calculate the integral in
Equation~(\ref{eq:convdef}) analytically for the functions at hand, the convolution was computed
numerically on the grid points of an $(r,z)$ grid with a spatial resolution of 10~pc and
$0 \leq r,z \leq 20$~kpc for this paper. SciPy's \citep{scipy} \texttt{tplquad} function in
Python~2.7 was used to directly evaluate the volume integrals at these points in the $y=0$ plane,
where $\tilde{B}_r = \tilde{B}_x$ and $B_y=\tilde{B}_y=0$. Therefore, the numerical smoothing
method introduced in this section serves as a fast and simple alternative to analytical approaches.

We compare the performance of the diffusive Galactic cosmic-ray propagation module in CRPropa~3.1
in the different field configurations in Section~\ref{sec:performance}. Bilinear interpolation of
the precomputed $\tilde{B}_r$ and $\tilde{B}_z$ values on the $(r,z)$ grid is used for the
implementation of the convolved field. While this interpolation routine suffices for the present
application in a propagation algorithm with a high grid resolution, for MHD simulations one should
instead choose a solenoidal interpolation routine based on, e.g., radial basis functions
\citep{radialbasis} or the vector potential \citep{vectorpotential}. \\

\begin{figure*}
  \centering
  \includegraphics[width=\textwidth]{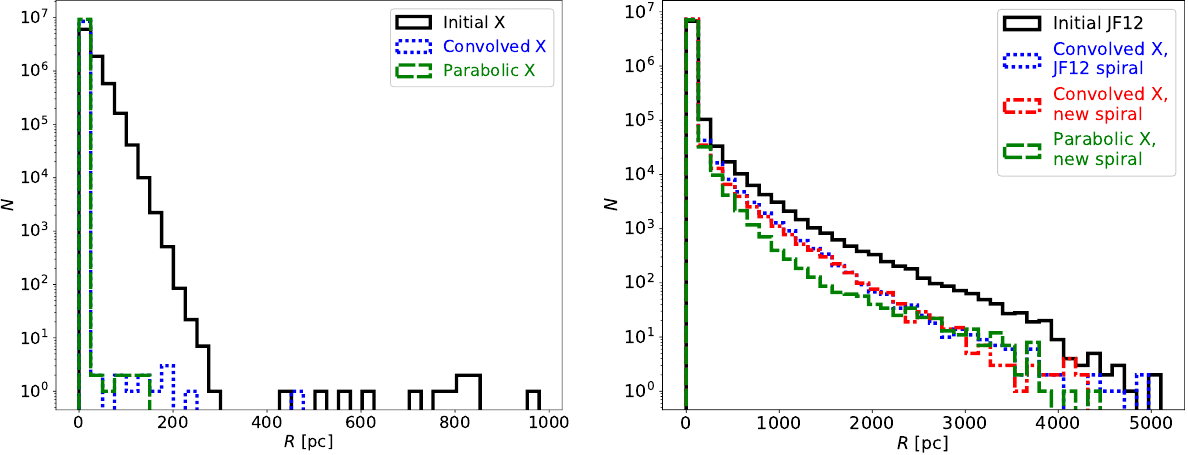}
  \caption{ \label{fig:crpropa}
    Histograms of field line deviations $R$ of pseudoparticles for purely parallel diffusion in
    different field configurations, comparing the initial JF12 field (black) with both the
    numerically convolved field (nonzero $\wx$, blue, as described in Section~\ref{sec:smoothing})
    and the X-field with parabolic insert (nonzero $\zs$, green, as described in
    Section~\ref{sec:parabolic}).
    Left: X-field only. Particles on average stay much closer to their starting field line if a
    smoothed field is used, as indicated by generally smaller values of $R$. Note also the large
    excursions exhibited by a relatively small number of ``outliers'' which are present for the
    standard JF12 field but are mostly absent from both smoothed fields.
    Right: The same for the total large-scale field, consisting of spiral disk, molecular ring,
    X-field, and toroidal halo. The red data was generated in a simulation combining a numerically
    convolved X-field with a modified (``new'') spiral field that uses flux redistribution at
    $r_1$ and a value of $r_2$ exceeding the extent of the computational domain, thus corresponding
    to the black curve in Figure~\ref{fig:p_delta}, while the standard JF12 spiral was used to
    obtain the blue data. Only particles whose deviation $R$ exceeds 0.05~pc are taken into
    account. Simulation parameters may be found in Section~\ref{sec:test_setup} and
    Table~\ref{tab:crpropa}. \\
  }
\end{figure*}

\begin{table*}[h!]
  \centering
  \caption{ \label{tab:crpropa}
    Results of the performance test simulations for different field configurations including the
    mean and median field line deviations of $N=10^7$ pseudoparticles with rigidity $\rho = 1$~PV,
    the mean call time of the \texttt{getField()} function in CRPropa, and the average time for a
    full test simulation. The presence of a factor of two between the smoothness parameters
    $w_{\rm X}$ and $\zs$ applied for this comparison is justified in Appendix~\ref{app:curvature}.
  }
  \begin{tabular}{l@{~}lr@{~}l|cccc}
    \multicolumn{2}{l}{Configuration} & \multicolumn{2}{l|}{Parameters [kpc]} & Mean Deviation $\left<R\right>$ [pc] & Median of $R$ [pc] & Call Time [$\upmu$s] & Simulation Time [s] \\ \hline
    \multicolumn{2}{l}{Unmodified X-field only} &     --      &  & $18.17 \pm 20.36$ & $11.61$ & $0.39$ & $44.2 \pm 1.1$ \\
    \multicolumn{2}{l}{Convolved X-field only}  & $\wx = 1.0$ &  & $ 0.22 \pm  0.29$ &  $0.16$ & $0.40$ & $37.0 \pm 1.2$ \\
    \multicolumn{2}{l}{Parabolic X-field only}  & $\zs = 0.5$ &  & $ 0.59 \pm  0.57$ &  $0.43$ & $0.41$ & $33.4 \pm 1.2$ \\
    \hline
    \multicolumn{2}{l}{Unmodified JF12 field} &      --     &  & $16.20 \pm 69.86$ &  $1.66$ & $0.51$ & $62.4 \pm 0.8$ \\
    Convolved X, & JF12 spiral & $\wx = 1.0$ &                 &  $6.53 \pm 43.19$ &  $0.45$ & $0.54$ & $62.9 \pm 0.9$ \\
    Convolved X, & new spiral  & $\wx = 1.0$ &; $\delta = 3.0$ &  $6.01 \pm 40.33$ &  $0.49$ & $0.77$ & $98.4 \pm 1.0$ \\
    Parabolic X, & new spiral  & $\zs = 0.5$ &; $\delta = 3.0$ &  $5.08 \pm 30.78$ &  $0.53$ & $0.74$ & $93.0 \pm 0.9$ \\
    \hline
  \end{tabular}
\end{table*}

\section{Performance Comparison in CRPropa}
\label{sec:performance}

\subsection{Test Setup}
\label{sec:test_setup}

Besides the avoidance of an infinitely strong current sheet, the modifications to the JF12 field
which were detailed above were also motivated by its application as the background field for
Galactic cosmic-ray propagation simulations. The publicly available \mbox{CRPropa 3.1} code
\citep[][see \url{https://crpropa.desy.de}]{AlvesBatista_EA:2016, Merten_EA:2017} was used for
testing the performance of the initial and modified JF12 fields in such applications. The
low-energy extension of this code (module ``DiffusionSDE'') is based on stochastic differential
equations and propagates individual phase-space elements with an anisotropic diffusion tensor,
such that the turbulent components of the GMF enter the simulation only implicitly. The algorithm
uses an adaptive 5(4)-Fehlberg algorithm with Cash--Karp coefficients \citep{Cash_EA:1990} in order
to determine the tangent vector to the magnetic field at each step via field line integration.
The tangent vector is then used to construct the local Frenet trihedron of the field line in which
the actual propagation step is performed. Since the algorithm relies on field line integrations
with adaptive step size, one may hope to reduce numerical errors and simulation time by
introducing smoother field lines with larger radii of curvature.

In order to quantitatively compare the accuracy of the field line integration for different field
configurations, the artificial test scenario of purely parallel diffusion with respect to the
magnetic field lines was considered. The numerical error of the simulation may then be assessed by
computing the spatial distance $R$ of the phase-space element position to its initial field line
after a given trajectory length. As cosmic rays experience not only deflections in the parallel
direction (along the magnetic field) but also perpendicular to it, such a simulation will most
likely not reflect reality. However, pure parallel diffusion can be seen as the computationally
most challenging limit for the field line integrator. Performing well in the case of pure parallel
diffusion will most likely also result in good (or even better) performance in other, less
idealized scenarios. For further discussions of realistic values of the ratio between parallel and
perpendicular diffusion coefficients, see e.g., \citet{Shalchi:2009} and references therein.

In these tests, a total of $N = 10^7$ pseudoparticles (``\mbox{CRPropa} candidates'') with a
rigidity of $\rho = 1$~PV were propagated diffusively on trajectories with a maximum total length
of 50~kpc. This particular rigidity was chosen as smaller rigidities lead to smaller step sizes
and better results, whereas the diffusive transport approximation may not be valid at larger
rigidities.
The injection of these candidates was carried out randomly at 2000 source positions, which were
uniformly distributed in a cylindrical volume with 1~kpc~$\leq r <$~15~kpc and $|z| < 300$~pc,
avoiding the central region within 1~kpc. For these source positions, field lines were generated
by second-order Heun integration with a fixed step size of 0.1~pc and a total length of 70~kpc.
Concerning the step sizes $r_{\text{min}}, r_{\text{max}}$ and relative error tolerance $\varepsilon$
of the adaptive propagation module, the values $r_{\text{min}} = 0.01$~pc, $r_{\text{max}} = 1$~kpc,
and $\varepsilon = 10^{-4}$ recommended by \citet{Merten_EA:2017} were used.  Finally, each
candidate was deactivated upon either reaching the maximum trajectory length, entering a region
without magnetic field, or leaving the simulation volume at a distance of 20~kpc from the origin.
Afterwards, the minimum distance $R$ between the particle's final position and the initial field
line was computed.

To ensure that field line integration via the Heun scheme is indeed able to generate nodes of the
``reference field lines'' that are sufficiently close to the analytical field lines which they are
to represent, the minimum distance computation in the ``X-field only'' test cases (see below) was
tentatively repeated by analytically computing the $\rp$ and $\rs$ labels of a particle's initial
and final positions in the initial JF12 and the parabolic X-field, respectively. The obtained
differences of the mean field line deviations were found to be in the milliparsec range, thus
justifying the use of the Heun method also for the full field, for which analytical field line
labels are not available. \\

\subsection{Results}
\label{sec:test_results}

Table~\ref{tab:crpropa} summarizes the parameters of performed tests and the respective
performance results regarding both field line deviation and runtime, while
Figure~\ref{fig:crpropa} displays the statistics of ``field line fidelity'' in each case.
The ad~hoc values for parameters $\zs$ and $\delta$ were chosen for the simulations to test
whether the introduction of these parameters is in principle able to improve the performance of
the propagation algorithm. These numerical tests need to be repeated once a new fit of the
modified JF12 model to observational data has been performed.

In the first set of tests, only the X-field was present. The left panel of Figure~\ref{fig:crpropa}
clearly shows that, while the majority of pseudoparticles stay relatively close to their
respective field lines, the original JF12 X-field also generates a small number of cases with
large excursions. It can also be seen that both smoothing methods are indeed able to eliminate
most of these outliers. Together with the higher degree of ``field line fidelity,'' as
indicated by generally much lower values of $R$, this demonstrates that the smoothing
achieves the desired effect, as anticipated.

The second set of simulations uses the full large-scale field including, in particular, the
toroidal halo and spiral disk field --~modulated in $z$ direction according to
Equation~(\ref{eq:Lminus1})~-- with components replaced according to the bottom part of
Table~\ref{tab:crpropa}. Our new spiral field was not closed at the outer 20~kpc boundary of the
simulation volume for these tests, but was allowed to extend unaltered up to the boundary of the
computational domain.
While the right panel of Figure~\ref{fig:crpropa} shows that the different smoothing methods did
not completely eliminate all outliers, the mean and median deviations given in
Table~\ref{tab:crpropa} indicate that the accuracy of the diffusion algorithm was improved in both
cases, with the analytic smoothing methods giving the best results for the total field.

We finally note that, when interpreting these results and in particular the magnitude of typical
$R$ values, it should be kept in mind that pseudoparticle trajectories are not to be considered
in isolation, and that the statistical weight of outliers, even those in the kiloparsec range,
will be rather small in any ensemble of reasonable size. 
Note also that the exact results presented here depend on the chosen diffusion coefficient.
A different choice of diffusion model or, e.g., rigidity of the pseudoparticles will certainly
change the numbers but will most likely leave the general shape of the distributions unaltered.

Concerning the runtime for simulations in the different field configurations, the pure call time of
the \texttt{getField()} function was evaluated $10^7$ times. It is no surprise that the modified
field calls take slightly longer as the evaluations are significantly more complicated than in the
initial JF12 field. However, one might hope that fewer refinements of the adaptive field line
integration step size are needed for smooth field lines, which could outweigh the call time
disadvantage. Consequently, full test simulations with a more realistic 0.1 ratio of perpendicular
to parallel diffusion (and all other settings as above) were conducted for the different field
setups using $N=10^6$ particles, and the average simulation time for five simulations was measured.
As can also be seen from Table~\ref{tab:crpropa}, the simulation times are indeed slightly reduced
for the smoothed X-fields compared to the original JF12 X-field. On the other hand, the simulation
runtime in the total field runs increased when a smoothed spiral field was used. In addition to
the increased function call times, this is possibly also caused by the introduction of new field
line kinks in the total field and the correspondingly increased number of subdivisions for the
computation of the modified curved trajectories.

Finally, we note that these analytical improvements to the GMF model are going to be available
with the latest version of the CRPropa software, at this time using parameters as given in
Table~\ref{tab:crpropa}. It can be used in the same way as the original implementation of the
field in the \texttt{JF12Field} module. \\

\section{Summary and Conclusions}
\label{sec:summary}

In this work, we propose, derive, and discuss two major modifications to the popular JF12 model
of the Galaxy's large-scale magnetic field. The first of these modifications consists of the
insertion of transition layers at the inner and outer rim of the spiral disk in which incoming
and outgoing magnetic field lines are redistributed, resulting in the spiral field now being fully
divergence-free also at its inner and outer boundary.
As a possible alternative to the latter, the disk field could also be continued outwards
indefinitely, thus avoiding an explicit flux closure by moving it to spatial infinity, but
incurring a possibly undesired excess in magnetic field energy.

The second, independent modification concerns the poloidal X-type field component and serves to
remove the sharp kinks of field lines which the latter exhibits at the Galactic midplane. These
kinks are either removed by a numerical convolution technique, or analytically replaced with
smooth parabolic inserts, which also fully satisfy the divergence constraint. As a minor issue,
we point out that the spherical cutout surrounding the origin can and should be removed to warrant
solenoidality also near the Galactic center. A simple way to ensure the X-field's solenoidality
also at large distances is to depart from the original model by allowing the field to continue
unaltered without explicit bounds. Although the very valid option of a dipolar field line closure
at finite distances and with a finite energy content -- now in the poloidal plane but otherwise in
line with what we suggest at the inner spiral disk rim -- exists as well, we refrain from a
further investigation of this possibility at this point.

Finally, we employ both smoothing techniques for a quantitative comparison in the framework of
numerical cosmic-ray particle tracing using the CRPropa framework, and demonstrate the particles'
superior field line fidelity of the modified X-type field over its unmodified predecessor.
A similar performance improvement could be found for the total field, with all suggested
modifications performing on a comparable level. We speculate that the observed slight superiority
of the analytical smoothing method might not necessarily prevail in other numerical settings.

In summary, we argue that, in addition to the observed performance improvement of the smoothed
X-field in our exemplary CRPropa test runs, this modified field also represents a useful option
for other applications, notably from the field of MHD simulations because it avoids an
unphysically strong current sheet in the Galactic plane. On the other hand, many applications
relying on a GMF may not at all be hampered by current sheets or kinking field lines. For those,
the original, unsmoothed X-field clearly continues to be the model of choice due to its
comparatively simpler form and ease of implementation. The divergence-free corrections of both
the spiral and X-field, however, are crucial for physically relevant applications of the JF12
field model, and for this reason we consider it to be of high importance that they be taken into
account in future studies.
Therefore, all the modifications proposed in this work act to further improve on the
usefulness and physical realism of the popular JF12 GMF model, which, however,
will only come to full fruition once the extended set of parameters has been readjusted to ensure
continued consistency with observational data. \\

\begin{figure}[b]
  \centering
  \includegraphics[width=0.38\textwidth]{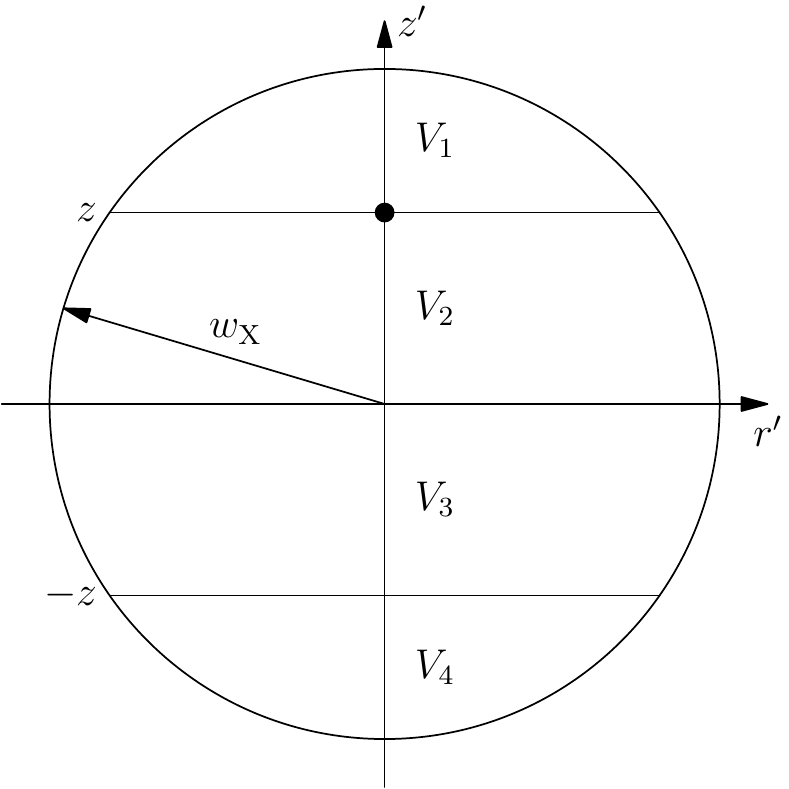}
  \caption{\label{fig:intsphere_sketch}
    Partition of the spherical integration volume $V$ into four vertically stacked subvolumes
    $V_{1...4}$ to ease computation of the convolution integral (\ref{eq:int_conv}). $B_r$ is
    negative for $z\p > z$, i.e., in $V_4$, and positive elsewhere. \\
  }
\end{figure}

\acknowledgments
\section*{Acknowledgments}

We are grateful to Horst Fichtner and the anonymous referee for valuable comments. Furthermore,
J.K.\ acknowledges financial support through the \textit{Ruhr Astroparticle and Plasma Physics (RAPP)
  Center}, funded as MERCUR project St-2014-040. \\

\appendix

\section{A. Matching Smoothing Parameters}
\label{app:curvature}

Since the analytical replacement method of Section~\ref{sec:parabolic} is very different from the
convolution method of Section~\ref{sec:smoothing}, a criterion is required that allows the
respective smoothing constants $\zs$ and $\wx$ to be chosen such that the resulting fields are of
comparable ``smoothness.'' To derive such a criterion, we employ the slightly simpler wedge-shaped
field
\begin{equation}
  B_z = 1, \quad B_r = \left\{ \begin{array}{rcl}
    s &:& z \ge 0 \\ -s &:& z < 0
  \end{array} \right. 
\end{equation}
with globally constant, rather than varying, inclination angle $\Tetn={\rm arccot}(s)$.
We consider the two smoothed versions of this field to be equivalent for the purpose of the
comparison detailed in Section~\ref{sec:performance} if their field lines have the same radius of
curvature at $z=0$. For the parabolic insertion method, this curvature radius follows directly
from Equation~(\ref{eq:fieldline_jf12}) as
\begin{equation}
  R_{\rm par} = \left( \left. \frac{\partial^2 \rF}{\partial z^2} \right|_{z=0} \right)^{-1}
  = \zs \tan \Tetn = \frac{\zs}{s} .
\end{equation}

Regarding the convolved field components $\tilde{B}_r$ and $\tilde{B}_z$, we first note that
$\tilde{B}_z = B_z = 1$, because $B_z$ is a global constant. The convolution formula
(\ref{eq:convdef}) for $B_r$ reads
\begin{equation}
  \label{eq:int_conv}
  \tilde{B}_r(z) = \int_V B_r(z-z\p) \, K \left(\sqrt{{r\p}^2+{z\p}^2}\right)
  2\pi r\p \dd r\p \dd z\p
\end{equation}
and the integration volume $V$ is a sphere of radius $\wx$ centered on $(r\p,z\p)=(0,0)$. Here, we
have implicitly set $r=0$ without loss of generality because $B_r$ is independent of $r$. Given
that we will eventually set $z$ equal to zero, we furthermore assume $z \in [0, \wx]$, also
without loss of generality.
As illustrated in Figure~\ref{fig:intsphere_sketch}, $V$ may be subdivided into four horizontally
sliced cutouts $V_{1...4}$ contained within the respective $z\p$ intervals $[-\wx, -z]$, $[-z, 0]$,
$[0, z]$, and $[z, \wx]$. Because $B_r = +s$ in $V_{1...3}$ and $B_r = -s$ in $V_4$, we see that
the contributions from $V_2$ and $V_3$ are equal, while those from $V_1$ and $V_4$ cancel. It is
therefore sufficient to perform the integration of Equation~(\ref{eq:int_conv}) just over
$V_3$ -- in which $B_r=s$ is a constant -- and then double the result. We may thus compute the
radius of curvature of the convolved field at $z=0$ according to
\begin{equation}
  \label{eq:getR_con}
  \begin{split}
    \frac{1}{R_{\rm con}} &= \pdv{}{z} \left. \left(
        \frac{\tilde{B}_r}{\tilde{B_z}}\right) \right|_{z=0}
    = \pdv{}{z} \left. \left(
        2 s \int_0^z \int_0^{\sqrt{\wx^2-{z\p}^2}}
        K \left(\sqrt{{z\p}^2+{r\p}^2} \right) \, 2\pi r\p \, \dd r\p \, \dd z\p
      \right) \right|_{z=0} \\
    &= 4\pi s \left. \int_0^{\sqrt{\wx^2-z^2}}
      K \left(\sqrt{z^2+{r\p}^2} \right) r\p \, \dd r\p \right|_{z=0}
    = 4\pi s \int_0^{\wx} K \left (r\p \right) r\p \, \dd r\p \\
    &=  4\pi s \, {\cal N} \int_0^1 \exp \left[ (u^2-1)^{-1} \right] \wx^2
    \, u \, \dd u = 4\pi s \, {\cal N} \wx^2 J_1
  \end{split}
\end{equation}
with the shorthand definition
\begin{equation}
  J_n \equiv \int_0^1 u^n \exp \left[ (u^2-1)^{-1} \right] \dd u .
\end{equation}
Inserting the normalization condition
\begin{equation}
  \frac{1}{\cal N} = \int_0^{\wx} K(r\p) \, 4\pi {r\p}^2 \dd r\p
  = 4\pi \int_0^1 \exp \left[ (u^2-1)^{-1}\right] \wx^3 \, u^2 \dd u
  = 4\pi \wx^3 J_2
\end{equation}
of kernel $K$ into Equation~(\ref{eq:getR_con}) leads us to $R_{\rm con} = (J_2/J_1) (\wx/s)$.
The condition $R_{\rm par} = R_{\rm con}$ is therefore equivalent to
\begin{equation}
  \frac{\wx}{\zs} = \frac{J_1}{J_2} \approx 2.114,
\end{equation}
independently of inclination angle. This justifies choosing parameters of ratio
$\wx / \zs = (1~{\rm kpc})/(500~{\rm pc}) = 2$ in Table~\ref{tab:crpropa}.

\vspace*{1cm}

\bibliographystyle{apj}
\bibliography{improve-JF12}

\end{document}